\documentclass[12pt]{article}
\usepackage{epsfig}
\usepackage{amssymb,amsmath}

\newcommand{\bea}{\begin{eqnarray}}
\newcommand{\eea}{\end{eqnarray}}

 \makeatletter
\def\alt{\mathrel{\mathpalette\gl@align<}}
\def\agt{\mathrel{\mathpalette\gl@align>}}
\def\gl@align#1#2{\lower.6ex\vbox{\baselineskip\z@skip\lineskip\z@
\ialign{$\m@th#1\hfil##\hfil$\crcr#2\crcr\sim\crcr}}} \makeatother

\begin{document}
\begin{flushright}
BA-08-22  \\
\end{flushright}
\vspace*{1.0cm}

\begin{center}
\baselineskip 20pt {\Large\bf NMSSM and Seesaw Physics at LHC }
\vspace{1cm}

{\large Ilia Gogoladze$^{a,}$\footnote{ E-mail:
ilia@physics.udel.edu\\ \hspace*{0.5cm} On  leave of absence from:
Andronikashvili Institute of Physics, GAS, 380077 Tbilisi, Georgia.}
, Nobuchika Okada$^{b,c,}$\footnote{ E-mail: okadan@post.kek.jp} and
Qaisar Shafi$^{a,}$ } \vspace{.5cm}

{\baselineskip 20pt \it
$^a$Bartol Research Institute, Department of Physics and Astronomy, \\
 University of Delaware, Newark, DE 19716, USA \\
\vspace{2mm}
$^b$Department of Physics, University of Maryland,
 College Park, MD 20742, USA  \\
\vspace{2mm}
$^c$
Theory Group, KEK, Tsukuba 305-0801, Japan }
\vspace{.5cm}

\vspace{1.5cm} {\bf Abstract}
\end{center}

We consider extensions of the next-to-minimal supersymmetric model
(NMSSM) in which the observed neutrino masses are described  in
terms of effective dimension six (or seven) rather than dimension
five operators. {All such operators respect the discrete symmetries
of the model.} The new particles associated with the double (or
triple) seesaw mechanism can have sizable couplings to the known
leptons, even with a TeV seesaw scale. In the latter case some of
these new short-lived particles could be produced and detected at
the LHC.

\thispagestyle{empty}

\newpage

\addtocounter{page}{-1}

\baselineskip 18pt

The next-to-minimal supersymmetric standard model (NMSSM)
\cite{NMSSM} provides a well motivated extension of MSSM in which
the $\mu$ problem of the latter is resolved through the introduction
of a gauge singlet  superfield $S$. This (NMSSM) extension has
several phenomenological consequences. For instance, the upper bound
on the lightest (CP-even) MSSM Higgs scalar can be increased from
around 125 GeV to close to 140 GeV \cite{Hmass}, and the little
hierarchy problem encountered in the  MSSM is ameliorated
\cite{LHprob1, LHprob2}. New Higgs boson decay channels into CP-odd
scalars appear \cite{LHprob2, Haa, Haa2} { which have an impact on
Higgs boson searches }, and there are new implications for dark
matter physics \cite{singlinoDM, RHsneutrino}.

In order to incorporate the observed solar and atmospheric neutrino
oscillations \cite{NuData}, we propose in this paper some extensions
of the NMSSM based on the seesaw mechanism for providing neutrino
masses. With the Large Hadron Collider (LHC) era about to unfold,
 we are especially interested in those seesaw extensions of the NMSSM
 which can be tested at the LHC. This leads us to consider dimension
 six and  seven (rather than dimension five)  operators for
the generation of the observed neutrino masses and  mixings. The
presence of the $S$ VEV in the NMSSM turns out to be an important
ingredient in implementing the double \cite{Dseesaw} (or triple)
seesaw mechanism. We consider several possibilities for
renormalizable models at high energies, which may include SU(2)
singlet, triplet and even additional doublet chiral superfields.

For the double or triple seesaw case, it is technically natural that
even with a seesaw mass scale as low as 1 TeV or so, the new
particles may have large couplings with the known leptons and Higgs
doublets. This  is generically not possible  for type I or III
seesaw. It is therefore an exciting possibility that some of the new
particles we introduce could be produced at the LHC and detected
through some distinctive decay signatures. We point out a way to
experimentally distinguish the new particles involved in
double seesaw from the ones in the conventional seesaw. We also
consider the low energy implications of lepton-number conserving but
lepton-flavor violating effective dimension six operators (in the
K\"ahler potential).

We begin by recalling the basic structure of the NMSSM \cite{NMSSM}.
We introduce a MSSM gauge singlet chiral superfield $S$ (with even
$Z_2$  matter parity) through the following superpotential terms:
\bea
 W \supset \lambda S H_u H_d +\frac{\kappa}{3} S^3,
\eea
where $\lambda$ and $\kappa$ are dimensionless constants, and $H_u$,
$H_d$ denote the MSSM Higgs doublets. A discrete $Z_3$ symmetry
under which S carries unit charge $\omega=e^{i 2 \pi/3}$ is
introduced in order to eliminate from $W$ terms that are linear and
quadratic in $S$, as well as the MSSM $\mu$ term. { Note that $S$
could be assigned a $Z_3$ charge $\omega^2$, but this leads to the
same dimension six and seven operators for neutrino masses. }  In
order to decide on the $Z_3$ charges of the MSSM Higgs doublets, we
require the presence in $W$ of Yukawa couplings at the
renormalizable level. There are {several} possible $Z_3$ charge
assignments for the matter superfields that are consistent with this
requirement as displayed in Table 1. { The $Z_3$ charge assignments
in Table 1 for the quark superfields is not unique, but this will
not be relevant for the discussion which follows.}

We assume that the breaking of supersymmetry   in the 'hidden'
sector induces electroweak scale soft scalar mass terms consistent
with $Z_3$ and $Z_2$:
\bea
 V_{\rm soft} =
  m_{H_u}^2 |H_u|^2 + m_{H_d}^2 |H_d|^2 + m_S^2 |S|^2
 + \left( m_\lambda S H_u H_d + m_\kappa S^3 +{\rm h.c.} \right).
\eea
The scalar component of $S$ acquires a non-zero VEV which generates
the desired MSSM $\mu$ term. The radiative electroweak breaking
scenario proceeds as in the MSSM case.

\begin{table}[t]
\begin{center}
\begin{tabular}{|c|c|c|c|c|c|c|c|c|c|}
\hline
& & $Q$ & $U^c$& $D^c$ & $L$ & $E^c$& $H_u$ &$H_d$ & $S$    \\
\hline
 case Ia & $Z_3$ & $\omega^2$ & 1 & 1 & $\omega^2$ & 1 & $\omega$& $\omega$& $\omega$ \\
 \hline
case Ib & $Z_3$ & $1$ & $\omega$ & $1$ & $\omega$ & $\omega^2$ & $\omega^2$& $1$& $\omega$ \\
 \hline
 case Ic & $Z_3$ & $1$ & 1 & $\omega$ & $1$ & $\omega$ & 1 & $\omega^2$& $\omega$ \\
 \hline
 \hline
 case IIa & $Z_3$ & 1 & $\omega^2$ & $\omega^2$ &  1& $\omega^2$ & $\omega$& $\omega$& $\omega$\\
 \hline
 case IIb & $Z_3$ & 1 & $\omega$ & $1$ & $\omega^2$ & $\omega$ & $\omega^2$& $1$& $\omega$\\
 \hline
 case IIc & $Z_3$ & 1 & $1$ & $\omega$ &  $\omega$& $1$ & $1$& $\omega^2$& $\omega$\\
 \hline
 \hline
 case IIIa & $Z_3$ & $\omega$ &  $\omega$ & $\omega$ & $\omega$ & $\omega$& $\omega$& $\omega$& $\omega$ \\
 \hline
 case IIIb & $Z_3$ & $1$ &  $\omega$ & $1$ & $1$ & $1$& $\omega^2$& $1$& $\omega$ \\
 \hline
 case IIIc & $Z_3$ & $1$ &  $1$ & $\omega$ & $\omega^2$ & $\omega^2$& $1$& $\omega^2$& $\omega$ \\
 \hline
\end{tabular}
\end{center}
\caption{ $Z_3$ charge assignments of the NMSSM  superfields
 which correspond to  dimension five, six or seven effective
 operators for neutrino masses.
} \label{tab:11}
\end{table}

According to the charge assignments in Table~1, neutrino masses
arise from  effective  dimension five, six or seven   operators:
\begin{eqnarray}
&&{\mbox{case Ia-Ic:}}~~~~~~~~~~~~ \frac{L L H_u H_u}{M_5},
 \label{sw1} \\
&&{\mbox{case IIa-IIc:}}~~~~~~~~~~ \frac{L L H_u H_u S}{M^2_6},
 \label{sw2} \\
&&{\mbox{case IIIa-IIIc:}}~~~~~~~~ \frac{L L H_u H_u S^2}{M^3_7},
 \label{sw3}
\end{eqnarray}
where $M_{5,6,7}$ denote the appropriate seesaw mass scales.

\begin{center}
{\large Case I}
\end{center}

Cases Ia-Ic correspond to the conventional dimensional five neutrino
operators. According to the three distinct ways to contract the
SU(2) indices in Eq.~(\ref{sw1}),
 there are three kinds of seesaw mechanisms:
 type I seesaw \cite{seesawI} mediated by the MSSM gauge singlet  fermions,
 type II seesaw \cite{seesawII} mediated by SU(2) triplet scalars
 with unit hypercharge, and type III seesaw \cite{seesawIII}
 mediated by SU(2) triplet fermions with zero hypercharge.
In this paper our focus will be mainly on dimension six
 and seven operators in Eq.~(\ref{sw2}) and Eq.~(\ref{sw3}).

\begin{center}
{\large Case II}
\end{center}

Following electroweak symmetry breaking, the dimension
 six operator (Eq.~(\ref{sw2})) induces  light neutrino Majorana mass given by
\bea
  m_\nu  \sim \left( \frac{v_u^2}{M} \right)
         \times \left( \frac{\langle S \rangle}{M} \right),
\eea
where we set $\langle H_u \rangle \equiv v_u $. Compared to the
conventional seesaw formula,
 we have an additional suppression factor $ {\langle S \rangle}/{M}$,
 so for this case the upper bound on the  seesaw mass scale is of order  $10^8$
GeV,
 assuming all Yukawa couplings   involved in the seesaw
 mechanism are of order unity. In practice, the seesaw scale can be
much lower.

 It is  interesting and instructive  to  propose an explicit model  which can
generate these  effective dimension six  operators.  We will show
that the masses of some of the new particles we introduce  can be
within reach of the LHC.

Note that case IIa-IIc give rise to the same dimension six operator
given by Eq.~(\ref{sw2}). The heavy fields which generate dimension
six operators for case IIa-IIc will differ only in the choice of
$Z_3$ charges. Thus,   we will consider  only case IIa,  which is
easily generalized for  IIb and  IIc.

As our first example  on how to generate dimension six operators
(Eq.~(\ref{sw2})), we introduce the following new particles in the
NMSSM (Figure~1),
 \bea
\begin{array}{c|cc|cc}
 \hspace{1cm}  & \mbox{SU(2)}   & \mbox{U(1)} & Z_3 & Z_2 \\
\hline
 N^c_j       & \bf{1}  &   0  &  \omega^2  & - \\
 N_j         & \bf{1}  &   0 &  \omega     & -
\end{array}
\label{typeI} \eea
where $i,j$ denote the generation indices. To reproduce the neutrino
oscillation data, we  need to introduce at least two generations
 of $N^c$ and $N$.

The  renormalizable superpotential terms involving only the new
chiral superfields is given by
\bea
 W \supset  Y_{ij} N^c_i (H_u L_j) + \frac{(\lambda_N)_{ij}}{2} S N_i N_j
     + m_{ij} N^c_i N_j.
\eea For $m_{ij}$  larger than the electroweak scale,
 we integrate out the heavy fields  $N^c_i$ and $N_i$
 under the SUSY vacuum conditions,
\bea
&&  \frac{\partial W}{\partial N}=0   \; \to \;
  N^c = - m^{-1} \lambda_N S N,  \nonumber \\
&&  \frac{\partial W}{\partial N^c}=0 \; \to \;
  N = - m^{-1} Y (H_u L),
\label{SUSYvac} \eea
where  the equations are to be understood in matrix form. Thus
 we arrive at the dimension six operator,
\bea
  W_{\rm eff}= \frac{1}{2}
     (H_u L)^T Y^T (m^{-1})^T  (\lambda_N S) m^{-1} Y  (H_u L).
\label{Eq-fig1}
\eea
For simplicity, we take $m_{ij} = M_6 \delta_{ij}$ and
 $(\lambda_N)_{ij}= \lambda_N \delta_{ij}$ .
Following the electroweak symmetry breaking,
 the neutrino Majorana mass matrix is generated:
\bea
  m_\nu = \frac{(Y^T Y) v_u^2}{M_6} \times
          \frac{\lambda_N \langle S \rangle}{M_6}.
\eea
This formula implies that even if $Y={\cal O}(1)$ and
 $M \sim 1$ TeV, the correct mass scale for the  light neutrinos
 can be reproduced by suitably adjusting  $\lambda_N$.

Note that the heavy fields integrated out also have an impact
  on the  K\"ahler potential.
Substituting  Eq.~(\ref{SUSYvac})
 into the canonical K\"ahler potential for the heavy fields,
 $\int d^4 \theta (N^\dagger N + N^{c \dagger} N^c)$,
 we obtain (see Figure~2)
\bea
 {\cal K}_{\rm eff} = (H_u L)^\dagger Y^\dagger (m^{-1})^\dagger m^{-1} Y (H_u L)
 + \cdots,
\label{Eq-fig2} \eea
where the ellipsis  denote  higher order terms. Following  the
electroweak symmetry breaking,
 this dimension six operator induces flavor-dependent
 corrections to the kinetic term
 of the left-handed neutrinos \cite{dim6I}.
Taking again  for simplicity  $m_{ij}= M_6 \delta_{ij}$ and
 $(\lambda_N)_{ij}= \lambda_N \delta_{ij}$,
 the modified  kinetic term of the left-handed neutrinos is found to be
\bea
 {\cal L}_{\rm kin} =
  i \overline{\nu_{L i}} \left(
  \delta_{ij}+\epsilon^N_{ij} \right)
 \gamma^\mu \partial_\mu \nu_{L j},
\eea
where
\bea
  \epsilon^N_{ij}= \frac{v_u^2}{M_6^2} (Y^\dagger Y)_{ij}.
\eea

In the presence of the flavor-dependent kinetic term,
 the relation between the mass ($\tilde{\nu})$ and flavor
 ($\nu$) eigenstates of the light Majorana neutrinos is given by
 $\nu_i  = ({\cal N} U_{\rm MNS})_{ij} \tilde{\nu}_j$,
 with $U_{MNS}$ the usual neutrino mixing matrix and
\bea
   {\cal N}_{ij} \simeq \delta_{ij}-\frac{1}{2} \epsilon^N_{ij},
\eea
in the approximation  $ \epsilon^N_{ij} \ll 1$. Note that the matrix
${\cal N}$ is not unitary. This  causes interesting modifications in
 both the charged and neutral currents (involving light neutrinos)
 in the SM:
\bea
 J_\mu^{CC} = \overline{ \tilde{e}_{Li}}
 \gamma_\mu {\cal N}_{ij} \tilde{\nu}_{L j}, \; \;
 J_\mu^{NC} = \frac{1}{2}
   \overline{\tilde{\nu}_{Li}} \gamma_\mu
   \left({\cal N}^\dagger {\cal N} \right)_{ij} \tilde{\nu}_{L j} ,
\eea
 expressed in terms of the mass eigenstates.
It turns out \cite{dim6II, dim6III} that
 the elements of $|{\cal N}{\cal N}^\dagger|$
 are somewhat severely constrained by
 the current experimental data on neutrino oscillations,
 $W$ and $Z$ boson decays,  and flavor-violating decays of leptons:
 $Y_{ij} \lesssim 0.1$ for $M_6 \sim 1$ TeV.
On the other hand, new signals of CP-violation related to  this
non-unitary leptonic mixing may be observed
 in future neutrino oscillation experiments \cite{nuoscillation}.

An alternative way to generate the effective dimension six
operator (Eq.~(\ref{sw2})) is to replace the singlet superfields
 in Eq.~(\ref{typeI}) with two SU(2) triplets with
 zero-hypercharge:
\bea
\begin{array}{c|cc|cc}
 \hspace{1cm}  & \mbox{SU(2)}   & \mbox{U(1)} & Z_3 & Z_2 \\
\hline
 \Delta^c_j  & \bf{3}  &   0  &  \omega^2  & - \\
 \Delta_j    & \bf{3}  &   0 &  \omega     & -
\end{array}
\label{typeIII}
\eea
The superpotential in this case is given by
\bea
 W \supset  Y_{ij} (H_u \Delta^c_i L_j)
   + \frac{(\lambda_{\Delta})_{ij}}{2} S \; {\rm tr}\left[\Delta_i \Delta_j \right]
   + m_{ij} {\rm tr}\left[ \Delta^c_i \Delta_j \right].
\eea
Integrating out the heavy triplets
 give rise to  dimension six operators in the  superpotential
 and the K\"ahler potential.
Substituting the various  VEVs,
 we obtain the  light neutrino Majorana masses and
 flavor-violating kinetic terms.
One difference from the previous (singlet) case is that
 dimension six operators are also  induced  for the charged
 leptons by integrating out the  heavy charged fields
 in the SU(2) triplets.
Again, in the mass basis, a non-unitary mixing matrix is
 induced, whose elements are constrained by the current
 experimental data \cite{dim6III}:
 $Y_{ij} \lesssim 0.01-0.1$ for $m_{ij} \sim 1$ TeV.

Yet another  way for generating the  dimension six operator  is to
introduce four additional  SU(2) triplets with unit hypercharge (see
Figure~3): \bea
\begin{array}{c|cc|cc}
 \hspace{1cm}  & \mbox{SU(2)}   & \mbox{U(1)} & Z_3 & Z_2 \\
\hline
 \Delta^c        & \bf{3} & +1 &  1 & + \\
 \bar{\Delta}^c  & \bf{3} & -1 &  1 & + \\
 \Delta          & \bf{3} & -1 & \omega & + \\
 \bar{\Delta}    & \bf{3} & +1 & \omega^2& + \\
\end{array}
\label{typeII} \eea
The additional contributions to the NMSSM superpotential in this
case contain the following terms
 \bea
 W \supset  Y_{ij} (L_i \Delta^c L_j) + Y_H (H_u \Delta H_u)
   + \lambda_N S \;
   {\rm tr}\left[ \bar{\Delta}^c \bar{\Delta} \right]
   + m_c \; {\rm tr}\left[ \bar{\Delta}^c  \Delta^c \right]
   + m \; {\rm tr}\left[ \bar{\Delta}    \Delta   \right].
   \label{hh2}
\eea
In this case, the effective K\"ahler potential
 after integrating out the heavy triplets
 is found to be of the form,
\bea
 {\cal K}_{\rm eff} \sim Y^\dagger_{ij} Y_{k l}
  \frac{L^\dagger_i L^\dagger_j L_k L_l}{m_c^2} ,
\label{Eq-fig3} \eea
such that  lepton-flavor violating four-Fermi interaction is
induced. The Yukawa coupling $Y_{ij}$ is constrained by
 the data from   flavor-violating decays of leptons \cite{dim6III}:
 $Y_{ij} \lesssim 0.01-0.1$ for $m_c \sim 1$ TeV.

As an example of a more elaborate model, let us introduce the
 following additions to the NMSSM:
\bea
\begin{array}{c|cc|cc}
 \hspace{1cm}  & \mbox{SU(2)}   & \mbox{U(1)} & Z_3 & Z_2 \\
\hline
 H_{u j}^\prime   & \bf{2} & +1/2 & \omega^2 & - \\
 H_{d j}^\prime   & \bf{2} & -1/2 & \omega   & - \\
  \Delta^c_j      & \bf{3} & 0 & \omega      & - \\
 \bar{\Delta}^c_j & \bf{3} & 0 & \omega^2    & - \\
\end{array}
\label{type4}
\eea
The relevant  part of the   superpotential is given by
\bea
 W \supset  Y_S^{ij} S (H_{u i}^\prime L_j)
    + (\lambda_\Delta)_{ij} (H_u \Delta^c_i H_{d i}^\prime)
    + Y_\Delta^{ij} (H_u \bar{\Delta}^c_i L_j)
    + m_H^{ij}       \; (H_{u i}^\prime H_{d j}^\prime)
    + m_\Delta^{ij}  \; {\rm tr} \left[\bar{\Delta}^c_i  \Delta^c_j \right].
\eea
The dimension six operators arise as shown in Figure~4.
The corresponding part of the K\"ahler potential takes
the form (symbolically), \bea
 {\cal K}_{\rm eff} \sim
 Y_S^\dagger Y_S \frac{L^\dagger L S^\dagger S}{m_H^2}
 + Y_\Delta^\dagger Y_\Delta \frac{L^\dagger L H_u^\dagger H_u}{m_\Delta^2},
\eea
and the flavor-dependent kinetic terms are generated
via the VEVs of $S$ and $H_u$.
Comparing to the constraints on $Y$s in the previous cases,
we read the current experimental bounds as
$Y_S^{ij}, Y_\Delta^{ij} \lesssim 0.01-0.1$
for $m_H, m_\Delta \sim 1$ TeV.

\begin{center}
{\large Case III}
\end{center}

Cases IIIa-IIIc correspond to dimension seven operators for neutrino
masses and mixings (see Eq.~(\ref{sw3})). We note that this operator
can be generated by integrating out the same heavy fields which we
introduced  for generating the dimensional six operator. The main
difference is in the $Z_3$ charge assignments. It is obvious that
there are more possibilities  to generate dimension seven operators
for neutrino masses compared to the dimensional six case.
 We will provide one example of how to  generate  such an operator.

For Case IIIa, we introduce the following new particles
 in the NMSSM spectrum (see Figure~5),
\bea
\begin{array}{c|cc|cc}
 \hspace{1cm}  & \mbox{SU(2)}   & \mbox{U(1)} & Z_3 & Z_2 \\
\hline
 N^c_j     & \bf{1}  &   0  &  \omega    & - \\
 N_j       & \bf{1}  &   0  &  \omega^2  & - \\
 N^0_j     & \bf{1}  &   0  &  1         & -
\end{array}
\label{typeIV} \eea
 where $i,j$ denote  the generation indices. To reproduce the neutrino
oscillation data, we  need to introduce at least two generations
 of $N^c, N^0$ and $N$.

The relevant part of the  renormalizable superpotential involving
only the new chiral superfields is given by
\bea
 W \supset  Y_{ij} N^c_i (H_u L_j) + (\lambda_N)_{ij} S N_i N^0_j
     + m_{ij} N^c_i N_j+  \frac{1}{2} m^{\prime}_{ij} N^0_i N^0_j.
 \label{fig55}
\eea
 For $m_{ij}$ and $m^{\prime}_{ij}$  larger than the electroweak scale,
 we integrate out the heavy $N^c_i,  N^0_i$ and $N_i$
 under the SUSY vacuum conditions. After eliminating the heavy fields, we arrive at
 the dimension seven operator of the form:
\bea
 W_{\rm eff}= - \frac{1}{2}
  (H_u L)^T Y^T (m^{-1})^T  (\lambda_N S)^T (m^\prime)^{-1}
 (\lambda_N S) m^{-1} Y  (H_u L).
\label{Eq-fig5}
\eea
For simplicity, we take $m_{ij} = m^{\prime}_{ij} =M_7\delta_{ij}$ and
 $(\lambda_N)_{ij}= \lambda_N \delta_{ij}$.
Following the electroweak symmetry breaking,
 the neutrino Majorana mass matrix is generated:
\bea
  m_\nu = \frac{(Y^T Y) v_u^2}{M_7} \times
          \frac{\lambda_N^T \lambda_N \langle S^2 \rangle}{M_7^2}.
\label{n77} \eea
We can see from this formula that the upper bound for seesaw scale
 is $M_7 \sim 10^6$ GeV, assuming all Yukawa coupling
 in Eq.~(\ref{Eq-fig5}) are ${\cal O}(1)$. It is clear from Eq.~(\ref{n77})
that by suitably adjusting the parameter $\lambda_N$, the seesaw
scale $M_7$
 can easily be lowered to the  TeV range. This opens up the exciting
possibility that some of the new particles in Eq.~(\ref{fig55}) can
have ${\cal O}(1)$ Yukawa couplings  with the known matter fields.

If some of the new particles generating the seesaw mechanism have
masses around 1 TeV, they could be produced in hadron colliders
\cite{LHC-Seesaw}. In type I seesaw, the heavy Majorana neutrino
productions at hadron colliders via { $W$-boson exchange
 and $WW$-fusion process \cite{LHCtypeI, LHCtypeI-2, Smirnov}
 have been investigated. }
Signatures for the Majorana neutrinos could be observed
 through their lepton-number violating decays leading to
 like-sign dilepton production \cite{LHCtypeI-2, Smirnov}.
However, the production cross section is normally small,
 because the heavy neutrinos dominantly consist of
 the singlet Majorana neutrinos and the couplings
 between the heavy neutrinos and the $W$-boson are
 suppressed by a factor $Y v_u/M_5$, the mixing between
 left-handed neutrinos and right-handed heavy neutrinos
 induced by the seesaw mechanism.
Agreement with  the  neutrino oscillation data requires that,
 $Y$ should at most be around $10^{-5}$, and so the mixing is very small
 for $M_5 \sim 1$ TeV.
Some fine-tuning for the Dirac Yukawa matrix is necessary
 to keep the mixing angle as large as possible
 while reproducing the neutrino oscillation data \cite{Smirnov}.

The situation for type II or III seesaw is different,
 since the new particles involved in the seesaw mechanism
 have  gauge couplings with the photon, $W$  and $Z$-boson.
Thus, these new particles could be produced at the LHC
 through  processes mediated by these gauge boson.
The signature of such particles in type II seesaw
 \cite{LHCtypeII} and in type III seesaw \cite{LHCtypeIII}
 have been investigated in detail.
For example, the doubly-charged scalar in type II seesaw,
 once produced, may provide a clean signature through its decay
 into a pair of  same sign charged leptons \cite{LHCtypeII}.
The pair production of the singly-charged fermions
 in type III seesaw and their (lepton-number and/or lepton-flavor
 violating) decays into leptons and $W$, $Z$-boson or Higgs boson
 could be discovered at the  LHC.
Note that if the seesaw scale is around 1 TeV,
 the Dirac Yukawa couplings are very small, say,
 $Y={\cal O}(10^{-5})$ to provide a light neutrino
 mass  $m_\nu ={\cal O}$(1 eV).
This fact implies that the new particle production
 is accompanied by an extra signature \cite{LHCtypeIII}:
 The lifetime of the produced particles is long enough
 for the their decay vertices to be detectably
 displaced from the primary production vertex.

The new particles included in the conventional seesaw models
 also appear in  models with the double (or triple) seesaw mechanism.
The collider phenomenology for these particles is analogous to the
one in the conventional seesaw models. However, there is a crucial
difference arising from  the structure of the double (or triple)
seesaw mechanism. As we have already noted, in the double (or
triple) seesaw mechanism
 we can reproduce the light neutrino Majorana masses
 while keeping the Dirac Yukawa couplings as large as possible,
 so long as  the couplings among the new particles are suitably small.
Therefore, the production cross section of the heavy neutrinos
 in type I seesaw can be sizable.
If the Dirac Yukawa couplings are much larger than the values
 expected in the conventional seesaw models with
 a  TeV seesaw scale, the lifetimes  of the new particles can be relatively  short.
For a low seesaw scale this could be a distinguishing feature
between the conventional seesaw
 and the double (or triple) seesaw models.
In addition, we have seen  that in  models with  double
 (or triple) seesaw, new SU(2) doublets,
 which do not  appear in the conventional seesaw models,
 can be  present.
The phenomenology of these new particles would be
 worth investigating.
For example, the charged scalar in the doublet superfield
 ($H_u^\prime$), once produced, can decay into  charged-leptons
 and the fermionic component of the singlet superfield $S$.

{ Recently, it has been pointed out \cite{RHsneutrino}
 that if the right-handed neutrinos have couplings with
 the singlet S in the NMSSM, the lightest right-handed sneutrino
 can be a viable  cold dark matter candidate
 through its coupling with the Higgs bosons.
Our double (or triple) seesaw mechanism with right-handed neutrinos
 shares the same structure for the right-handed sneutrinos,
 so that this scenario can also work in our models.
}

Finally, let us recall that the presence of low scale seesaw can
alter the predictions for the SM Higgs boson mass
\cite{Gogoladze:2008gf}. In the NMSSM case, the coupling $Y_H (H_u
\Delta H_u)$ in  Eq.~(\ref{hh2}) will generate a tree level
contribution to the lightest CP-even Higgs boson mass
\cite{Espinosa:1998re}:
 \bea
 m_h^2=(m_h^2)_{NMSSM} + 4 Y_H^2 v_u^2 \sin^2 \beta,
 \label{hh3} \eea
where $\tan \beta =v_u/v_d$ and $(m_h^2)_{NMSSM}$ denotes the
standard NMSSM contribution.  There is a  constraint on the triplet
Higgs VEV arising from a global fit of electroweak data
\cite{delAguila:2008ks}:
 \bea
v_{\Delta} \lesssim 2 \; {\mbox{GeV}}.
 \label{hh4} \eea
If one assumes  that the triplet Higgs mass is a few TeV,
 the trilinear coupling $Y_H$ can be order unity.
In this case the  tree level contribution can help to make the Higgs
mass as heavy as 200 GeV (for $\tan \beta > 10$).  For a  more
precise  estimate, loop correction arising from the  Higgs triplet
should be included.

In summary, we have proposed extensions of the NMSSM particle
spectrum such that the observed neutrino masses are described in
terms of effective dimension six or seven (instead of dimension
five) operators. In addition to the models corresponding to type I,
II and III seesaw, we introduce more exotic possibilities with SU(2)
doublet superfields which do not appear in the conventional seesaw
models. The new heavy states  responsible for the seesaw mechanism
can have sizable couplings to the known leptons and may be detected
at the LHC. The low energy implications of lepton flavor violating
dimension six operators in the K\"ahler potential are briefly
discussed.

\section*{Acknowledgments}
N.O. would like to thank the Particle Theory Group of the University
of Delaware for hospitality during his visit. He would also like to
thank the Maryland Center for Fundamental Physics, and especially
Rabindra N. Mohapatra for their hospitality and financial support
during his stay. This work is supported in part by the DOE Grant \#
DE-FG02-91ER40626 (I.G. and Q.S.), GNSF grant 07\_462\_4-270 (I.G.),
{ the National Science Foundation Grant No. PHY-0652363 (N.O.)}, and
the Grant-in-Aid for Scientific Research from the Ministry of
Education, Science and Culture of Japan, \#18740170 (N.O.).


\begin{figure}[b]\begin{center}
\begin{picture}(160,200)
\put(-70,350){\begin{minipage}{8cm}
\unitlength=0.15mm
\begin{picture}(1000,1000)(-350,200)
\put(-350, 0){\line(1,0){700}}
\put(-250,0){\line(0,1){150}}
\put(   0,0){\line(0,1){150}}
\put( 250,0){\line(0,1){150}}
\put(-125,0){\circle*{20}}
\put(+125,0){\circle*{20}}
\put(-50,-90){\makebox(100,100){\Large{$\lambda_N$}}}
\put(-175,-100){\makebox(100,100){\Large{$m$}}}
\put(+75,-100){\makebox(100,100){\Large{$m$}}}
\put(-300,-100){\makebox(100,100){\Large{$Y$}}}
\put(+200,-100){\makebox(100,100){\Large{$Y$}}}
\put(-50, 140){\makebox(100,100){\Large{$S$}}}
\put(-300,140){\makebox(100,100){\Large{$H_u$}}}
\put(+200,140){\makebox(100,100){\Large{$H_u$}}}
\put(+150,-20){\makebox(100,100){\Large{$N^c$}}}
\put(+20,-20){\makebox(100,100){\Large{$N$}}}
\put(-230,-20){\makebox(100,100){\Large{$N^c$}}}
\put(-100,-20){\makebox(100,100){\Large{$N$}}}
\put(-430,-50){\makebox(100,100){\Large{$L$}}}
\put(320,-50){\makebox(100,100){\Large{$L$}}}
\end{picture}
\end{minipage}}
\end{picture}
\caption{ Supergraph leading to dimension six operator
 in the superpotential of Eq.~(\ref{Eq-fig1}).
}
\end{center}
\end{figure}
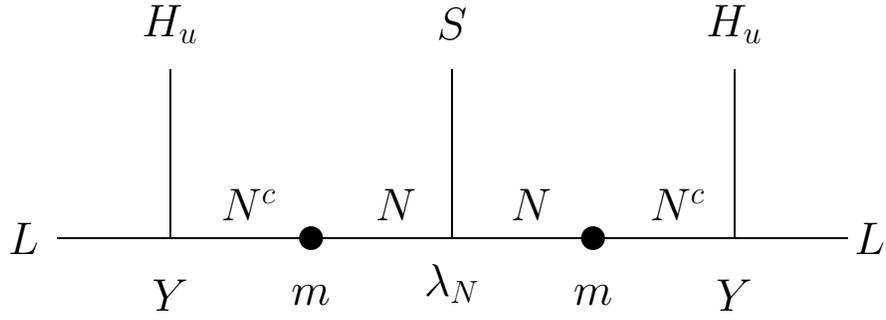
%

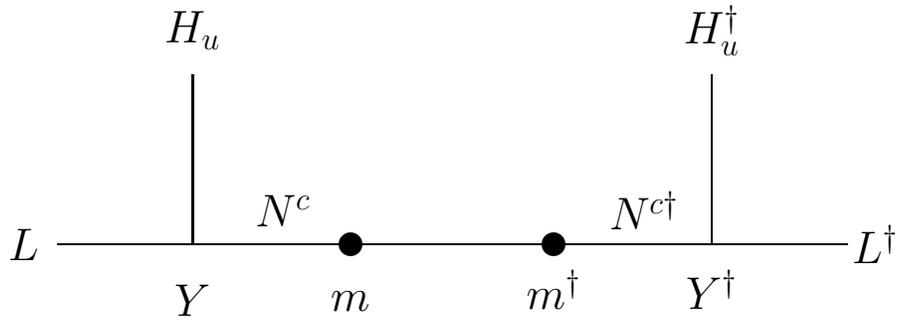
\begin{figure}[t]\begin{center}
\begin{picture}(160,200)
\put(-70,350){\begin{minipage}{8cm}
\unitlength=0.15mm
\begin{picture}(1000,1000)(-350,200)
\put(-350, 0){\line(1,0){700}}
\put(-230,0){\line(0,1){150}}
\put( 230,0){\line(0,1){150}}
\put(-90,0){\circle*{20}}
\put(+90,0){\circle*{20}}
\put(-140,-100){\makebox(100,100){\Large{$m$}}}
\put(+40,-90){\makebox(100,100){\Large{$m^\dagger$}}}
\put(-280,-100){\makebox(100,100){\Large{$Y$}}}
\put(+180,-90){\makebox(100,100){\Large{$Y^\dagger$}}}
\put(-280,140){\makebox(100,100){\Large{$H_u$}}}
\put(+180,140){\makebox(100,100){\Large{$H_u^\dagger$}}}
\put(+120,-20){\makebox(100,100){\Large{$N^{c \dagger}$}}}
\put(-200,-20){\makebox(100,100){\Large{$N^c$}}}
\put(-430,-50){\makebox(100,100){\Large{$L$}}}
\put(325,-50){\makebox(100,100){\Large{$L^\dagger$}}}
\end{picture}
\end{minipage}}
\end{picture}
\caption{ Supergraph leading to dimension six operator
 in the K\"ahler potential of Eq.~(\ref{Eq-fig2}).
}
\end{center}
\end{figure}
%

\begin{figure}[t]\begin{center}
\begin{picture}(160,200)
\put(-70,350){\begin{minipage}{8cm}
\unitlength=0.15mm
\begin{picture}(1000,1000)(-350,200)
\put(-350, 0){\line(1,0){700}}
\put(-250,0){\line(0,1){150}}
\put(   0,0){\line(0,1){150}}
\put( 250,0){\line(0,1){150}}
\put(-125,0){\circle*{20}}
\put(+125,0){\circle*{20}}
\put(-50,-90){\makebox(100,100){\Large{$\lambda_N$}}}
\put(-175,-100){\makebox(100,100){\Large{$m_c$}}}
\put(+75,-100){\makebox(100,100){\Large{$m$}}}
\put(-300,-100){\makebox(100,100){\Large{$Y$}}}
\put(+190,-100){\makebox(100,100){\Large{$Y_H$}}}
\put(-50, 140){\makebox(100,100){\Large{$S$}}}
\put(-300,140){\makebox(100,100){\Large{$L$}}}
\put(+200,140){\makebox(100,100){\Large{$H_u$}}}
\put(+140,-20){\makebox(100,100){\Large{$\Delta$}}}
\put(+15,-20){\makebox(100,100){\Large{$\bar{\Delta}$}}}
\put(-230,-20){\makebox(100,100){\Large{$\Delta^c$}}}
\put(-100,-20){\makebox(100,100){\Large{$\bar{\Delta}^c$}}}
\put(-430,-50){\makebox(100,100){\Large{$L$}}}
\put(320,-50){\makebox(100,100){\Large{$H_u$}}}
\end{picture}
\end{minipage}}
\end{picture}
\caption{ Supergraph leading to dimension six operator
 after integrating out the heavy triplets.
}
\end{center}
\end{figure}
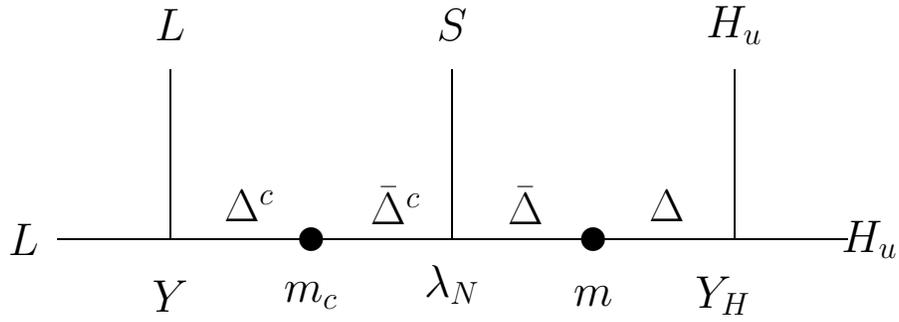

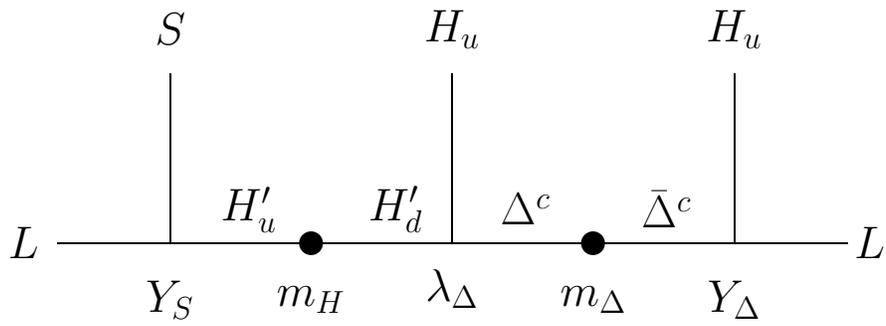
\begin{figure}[t]\begin{center}
\begin{picture}(160,200)
\put(-70,350){\begin{minipage}{8cm}
\unitlength=0.15mm
\begin{picture}(1000,1000)(-350,200)
\put(-350, 0){\line(1,0){700}}
\put(-250,0){\line(0,1){150}}
\put(   0,0){\line(0,1){150}}
\put( 250,0){\line(0,1){150}}
\put(-125,0){\circle*{20}}
\put(+125,0){\circle*{20}}
\put(-50,-90){\makebox(100,100){\Large{$\lambda_\Delta$}}}
\put(-175,-100){\makebox(100,100){\Large{$m_H$}}}
\put(+75,-100){\makebox(100,100){\Large{$m_\Delta$}}}
\put(-300,-100){\makebox(100,100){\Large{$Y_S$}}}
\put(+200,-100){\makebox(100,100){\Large{$Y_\Delta$}}}
\put(-50, 140){\makebox(100,100){\Large{$H_u$}}}
\put(-300,140){\makebox(100,100){\Large{$S$}}}
\put(+200,140){\makebox(100,100){\Large{$H_u$}}}
\put(+140,-20){\makebox(100,100){\Large{$\bar{\Delta}^c$}}}
\put(+15,-20){\makebox(100,100){\Large{$\Delta^c$}}}
\put(-230,-20){\makebox(100,100){\Large{$H_u^\prime$}}}
\put(-100,-20){\makebox(100,100){\Large{$H_d^\prime$}}}
\put(-430,-50){\makebox(100,100){\Large{$L$}}}
\put(320,-50){\makebox(100,100){\Large{$L$}}}
\end{picture}
\end{minipage}}
\end{picture}
\caption{ Supergraph with additional doublets leading to dimension
six operator. }
\end{center}
\end{figure}

\begin{figure}[t]\begin{center}
\begin{picture}(160,200)
\put(-70,350){\begin{minipage}{8cm} \unitlength=0.15mm
\begin{picture}(1000,1000)(-450,200)
\put(-600, 0){\line(1,0){700}}
 \put(-450, 0){\line(1,0){700}}
 \put(-350, 0){\line(1,0){700}}
\put(-500,0){\line(0,1){150}}
 \put(-250,0){\line(0,1){150}} \put(
0,0){\line(0,1){150}}
 \put(250,0){\line(0,1){150}}
\put(-375,0){\circle*{20}}
 \put(-125,0){\circle*{20}}
 \put(+125,0){\circle*{20}}
\put(-50,-90){\makebox(100,100){\Large{$\lambda_N$}}}
\put(-300,-90){\makebox(100,100){\Large{$\lambda_N$}}}
\put(-175,-95){\makebox(100,100){\Large{$m^\prime$}}}
\put(-430,-100){\makebox(100,100){\Large{$m$}}}
\put(+75,-100){\makebox(100,100){\Large{$m$}}}
\put(-560,-100){\makebox(100,100){\Large{$Y$}}}
\put(+200,-100){\makebox(100,100){\Large{$Y$}}}
\put(-50, 140){\makebox(100,100){\Large{$S$}}}
\put(-300,140){\makebox(100,100){\Large{$S$}}}
\put(-535,140){\makebox(100,100){\Large{$H_u$}}}
\put(+200,140){\makebox(100,100){\Large{$H_u$}}}
\put(-500,-20){\makebox(100,100){\Large{$N^c$}}}
\put(-350,-20){\makebox(100,100){\Large{$N$}}}
\put(+150,-20){\makebox(100,100){\Large{$N^c$}}}
\put(+20,-20){\makebox(100,100){\Large{$N$}}}
\put(-230,-20){\makebox(100,100){\Large{$N^0$}}}
\put(-100,-20){\makebox(100,100){\Large{$N^0$}}}
\put(-670,-50){\makebox(100,100){\Large{$L$}}}
\put(320,-50){\makebox(100,100){\Large{$L$}}}
\end{picture}
\end{minipage}}
\end{picture}
\caption{ Supergraph leading to dimension seven operator
 in the superpotential of Eq.~(\ref{Eq-fig5}).
}
\end{center}
\end{figure}
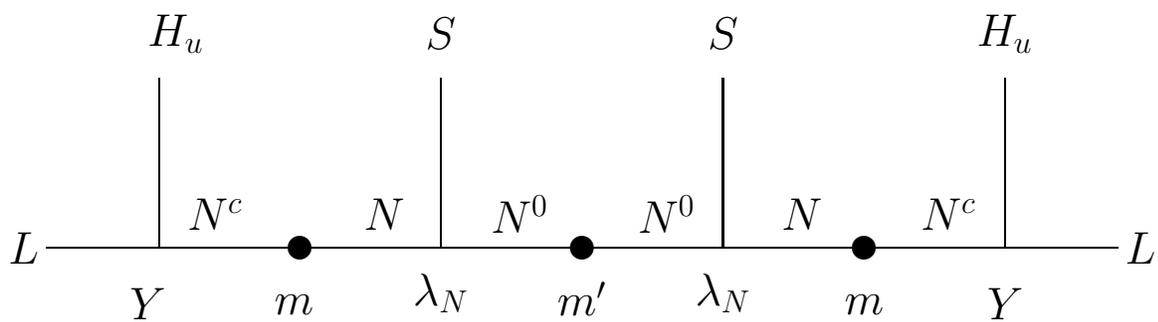

\end{document}